\documentstyle[onecolumn,psfig]{mn}

\newif\ifAMStwofonts

\def\mps{MPS model }
\def\mpsd{MPS model. }
\def\mpsc{MPS model, }
\def\delc{\delta_{\rm c}}
\def\nbody{{$N$}-body }
\newcommand{\der}{{\rm d}}
\def\delm{\Delta_{\rm m}}

\ifoldfss
  \ifCUPmtlplainloaded \else
    \NewTextAlphabet{textbfit} {cmbxti10} {}
    \NewTextAlphabet{textbfss} {cmssbx10} {}
    \NewMathAlphabet{mathbfit} {cmbxti10} {} 
    \NewMathAlphabet{mathbfss} {cmssbx10} {} 
  \fi
  \ifAMStwofonts
    \ifCUPmtlplainloaded \else
      \NewSymbolFont{upmath} {eurm10}
      \NewSymbolFont{AMSa} {msam10}
      \NewMathSymbol{\upi}     {0}{upmath}{19}
      \NewMathSymbol{\umu}     {0}{upmath}{16}
      \NewMathSymbol{\upartial}{0}{upmath}{40}
      \NewMathSymbol{\leqslant}{3}{AMSa}{36}
      \NewMathSymbol{\geqslant}{3}{AMSa}{3E}

       \let\le=\leqslant
       
    \fi
  \fi
\fi 

\ifnfssone
  \newmathalphabet{\mathit}
  \addtoversion{normal}{\mathit}{cmr}{m}{it}
  \addtoversion{bold}{\mathit}{cmr}{bx}{it}
  \newmathalphabet{\mathbfit} 
  \addtoversion{normal}{\mathbfit}{cmr}{bx}{it}
  \addtoversion{bold}{\mathbfit}{cmr}{bx}{it}
  \newmathalphabet{\mathbfss} 
  \addtoversion{normal}{\mathbfss}{cmss}{bx}{n}
  \addtoversion{bold}{\mathbfss}{cmss}{bx}{n}
  \ifAMStwofonts
    \ifCUPmtlplainloaded \else
      \UseAMStwoboldmath
      \makeatletter
      \new@mathgroup\upmath@group
      \define@mathgroup\mv@normal\upmath@group{eur}{m}{n}
      \define@mathgroup\mv@bold\upmath@group{eur}{b}{n}
      \edef\UPM{\hexnumber\upmath@group}
      \new@mathgroup\amsa@group
      \define@mathgroup\mv@normal\amsa@group{msa}{m}{n}
      \define@mathgroup\mv@bold\amsa@group{msa}{m}{n}
      \edef\AMSa{\hexnumber\amsa@group}
      \makeatother
      \mathchardef\upi="0\UPM19
      \mathchardef\umu="0\UPM16
      \mathchardef\upartial="0\UPM40
      \mathchardef\leqslant="3\AMSa36
      \mathchardef\geqslant="3\AMSa3E

       \let\le=\leqslant

    \fi
  \fi
\fi 

\ifnfsstwo
  \DeclareMathAlphabet{\mathbfit}{OT1}{cmr}{bx}{it}
  \SetMathAlphabet\mathbfit{bold}{OT1}{cmr}{bx}{it}
  \DeclareMathAlphabet{\mathbfss}{OT1}{cmss}{bx}{n}

  \SetMathAlphabet\mathbfss{bold}{OT1}{cmss}{bx}{n}
  \ifAMStwofonts
    \ifCUPmtlplainloaded \else
      \DeclareSymbolFont{UPM}{U}{eur}{m}{n}
      \SetSymbolFont{UPM}{bold}{U}{eur}{b}{n}
      \DeclareSymbolFont{AMSa}{U}{msa}{m}{n}
      \DeclareMathSymbol{\upi}{0}{UPM}{"19}
      \DeclareMathSymbol{\umu}{0}{UPM}{"16}
      \DeclareMathSymbol{\upartial}{0}{UPM}{"40}
      \DeclareMathSymbol{\leqslant}{3}{AMSa}{"36}
      \DeclareMathSymbol{\geqslant}{3}{AMSa}{"3E}

       \let\le=\leqslant

    \fi
  \fi
\fi 

\ifCUPmtlplainloaded \else
  \ifAMStwofonts \else 
    \def\upi{\pi}
    \def\umu{\mu}
    \def\upartial{\partial}
  \fi
\fi

\title[Testing the Modified PS Model]{Testing the Modified 
Press--Schechter Model against N-Body Simulations}

\author[Raig, Gonz\'alez-Casado \& Salvador-Sol\'e]{Andreu Raig$^{1}$, 
Guillermo Gonz\'alez-Casado$^{2}$ and Eduard Salvador-Sol\'e$^{1}$\\
$^{1}$Departament d'Astronomia i Meteorologia, Universitat de
Barcelona, Mart{\'\i} Franqu\`es 1, E-08028 Barcelona, Spain \\
$^{2}$Departament de Matem\`atica Aplicada II, Universitat 
Polit\`ecnica de 
Catalunya, Pau Gargallo 5, Edifici U, E-08028 Barcelona, Spain
}

\begin{document}

\maketitle

\begin{abstract}
\noindent A modified version of the extended Press--Schechter model
for the growth of dark-matter haloes was introduced in two previous
papers with the aim at explaining the mass-density relation shown by
haloes in high-resolution cosmological simulations. In this model
major mergers are well separated from accretion, thereby allowing a
natural definition of halo formation and destruction. This makes it
possible to derive analytic expressions for halo formation and
destruction rates, the mass accretion rate, and the probability
distribution functions of halo formation times and progenitor
masses. The stochastic merger histories of haloes can be readily
derived and easily incorporated into semi-analytical models of galaxy
formation, thus avoiding the usual problems encountered in the
construction of Monte Carlo merger trees from the original extended
Press--Schechter formalism. Here we show that the predictions of the
modified Press--Schechter model are in good agreement with the results
of $N$-body simulations for several scale-free cosmologies.
\end{abstract}


\section{INTRODUCTION}\label{intro}

The last decade saw the significant development of semi-analytical
techniques used to model galaxy formation in hierarchical cosmologies
(e.g., Lacey et al.~1993; Kauffmann, White \& Guiderdoni 1993; Cole et
al.~1994; Nulsen \& Fabian 1997; Avila-Reese \& Firmani 1998;
Cavaliere, Menci \& Tozzi 1999; Somerville \& Primack 1999; Cole et
al.~2000). The prediction of the merger histories of dark-matter haloes is
fundamental to all these techniques. Some recent schemes extract halo
evolution directly from cosmological \nbody simulations (Roukema et
al.~1997; van Kampen, Jim\'enez \& Peacock 1999; Kauffmann et
al.~1999). However, the most popular semi-analytical models (SAMs) use
Monte Carlo simulations of halo merger histories, the so-called merger
trees (e.g., Kauffmann et al.~1993; Somerville \& Primack 1999; Cole
et al.~2000) derived from the framework of the extended Press--Schechter
(EPS) theory (Bower 1991; Bond et al.~1991; Lacey \& Cole 1993,
hereafter LC93).

Despite its simplicity, the EPS theory has proved to be a very powerful
tool to model the merger histories of dark-matter haloes. The
predicted halo mass function is consistent with the results of \nbody
simulations. Although there are significant discrepancies at some
redshifts (e.g., Gross et al.~1998; Tormen 1998; Somerville et
al.~2000), they can be conveniently dealt with by means of different
corrections (Lee \& Shandarin 1998; Sheth \& Tormen 1999; Jenkins et
al.~2001; Sheth, Mo \& Tormen 2001). Furthermore, detailed
comparisons with \nbody simulations indicate that the EPS theory
provides a reliable statistical description of halo evolution
(Kauffmann \& White 1993; Lacey \& Cole 1994, hereafter LC94;
Somerville et al.~2000).

However, several issues concerning the practical implementation of
merger trees from the EPS theory require further study (see Somerville
\& Kolatt 1999; Somerville et al.~2000; Cole et al.~2000):

\begin{enumerate}
 \item The number (although not the mass) of progenitors of a halo
       diverges for progenitor masses approaching zero. Therefore
       it is necessary to introduce a minimum mass cut-off, or mass
       resolution, when constructing Monte Carlo merger
       trees to avoid infinite ramification. The
       problem is then how to account in a self-consistent way for the
       mass enclosed within unresolved haloes.

 \item To select the masses of resolved progenitors, one uses the
       conditional probability (provided by the EPS model) that a halo
       of a given mass at a given time has a progenitor of a
       smaller mass in an earlier epoch. However, this probability
       corresponds to {\it one single object} which implies that, having
       found one progenitor of a given mass, it cannot be
       used to find the mass of another progenitor. This introduces
       great uncertainties, namely, there is no way to decide how many
       progenitors above the resolution to consider nor the
       total mass that must be assigned to them.

 \item In merger trees, the branching of haloes into their resolved
       progenitors is imposed in a set of individual times which are
       separated by an arbitrary step, i.e., there is also some time
       resolution.  Progenitor mergers are then identified
       with these time-sparse merger nodes while they actually take
       place at some unknown moment between them.
\end{enumerate}       

Several ways have been proposed to enforce mass conservation in merger
nodes and simultaneously reproduce the conditional mass function of
the EPS theory (Kauffmann \& White 1993; LC93; LC94; Kitayama \& Suto
1996; Somerville \& Kolatt 1999; Somerville et al.~2000; Cole et
al.~2000). However, no fully satisfactory solution has been found to
date. For example, recent methods use non-arbitrary time steps
(Somerville \& Kolatt 1999; Cole et al.~2000), but the caveat always
remains that true mergers do not take place at the merger nodes where
progenitors are identified. In fact, the latter problem will always be
found in the EPS model because of its lack of a clear characterization
of halo formation and destruction.

LC93 introduced a prescription to characterize the formation and
destruction of haloes in the EPS model. The formation time is defined as
the time when half the halo mass is assembled into a single
progenitor. Likewise, the destruction time is defined as the time when the
halo doubles in mass. However, such definitions do not account for
the way mass is assembled. No distinction is made, for example,
between a halo which has increased its mass through one single capture
of another similar massive halo and one which has increased its mass
through consecutive captures of very small relative mass haloes. A
more realistic characterization of the formation and destruction of
haloes should account for this distinction. Only major mergers are
expected to produce, through violent relaxation, a substantial
rearrangement of the structure of haloes. Consequently, major mergers
can be regarded as causing the formation of new haloes with {\it
distinct} structural properties from the original ones which are
destroyed in the event. In contrast, minor mergers do not produce
significant disturbances in the equilibrium state and the structure of
the massive partner experiencing them. They simply cause a smooth mass
increase, which is commonly called accretion. Hence, in minor
mergers, the identity of the capturing halo with {\it essentially
unaltered} structural properties prevails.

Salvador-Sol\'e, Solanes \& Manrique (1998, hereafter SSM) introduced
the minimal self-consistent modification of the EPS model which
incorporates the previous natural definitions of halo formation and
destruction. In this new model, hereafter referred to as the modified
Press--Schechter (MPS) model, the distinction between major and minor
mergers is made through a phenomenological frontier $\delm$ in the
fractional mass captured by a halo. Captures above this threshold are
taken to be major mergers, while those below are considered minor
mergers which contribute to accretion. SSM found that, given an
effective value of $\delm$ of about $0.5$--$0.7$, the empirical
mass-density relation shown by haloes in high resolution \nbody
simulations of distinct hierarchical cosmologies (Navarro, Frenk \&
White 1997, hereafter NFW) is naturally reproduced under the
assumption proposed by NFW that the characteristic density of haloes
is proportional to the critical density of the universe at the time of
their formation. Raig, Gonz\'alez-Casado \& Salvador-Sol\'e (1998,
hereafter RGS) subsequently showed that the mass-density relation is,
in this case, not only consistent with, but actually implied by the
\mpsd To prove this, RGS assumed haloes endowed with a universal
density profile \`a la NFW. However, the same conclusion holds
whatever the form of the halo density profile (Gonz\'alez-Casado, Raig
\& Salvador-Sol\'e 1999). More importantly, using the \mpsc Manrique,
Salvador-Sol\'e \& Raig (2001) have recently shown that the true shape
\`a la NFW of halo density profiles is the natural consequence of
hierarchical clustering.

Hence, the \mps provides a consistent description of both the growth
history and the internal structure of haloes. Furthermore, as shown
here, the \mps overcomes the problems encountered with Monte Carlo
merger trees built from the EPS model (see also Salvador-Sol\'e et
al.~2001). The ability of the \mps to account for the structure of
dark-matter haloes strongly supports its validity. However, before
incorporation into a SAM of galaxy formation the \mps should be
thoroughly tested against cosmological \nbody simulations, as was the
EPS model (e.g., LC94). This is the aim of the present paper. In
particular, we propose to check the correct behaviour of all
quantities predicted by the \mps that represent a novelty over the
original EPS model, namely, the rates of halo formation and
destruction, the mass accretion rate, and the probability distribution
functions (PDFs) of halo formation times and progenitor masses.

In Section \ref{anmod} we review the \mpsd In Section \ref{nbsim} we
describe the \nbody simulations used to test the model and the way the
empirical quantities have been extracted from them. The comparison
between theory and simulations is presented in Section \ref{resul} and
the main results of this comparison summarized in Section \ref{diss}.

\section{THE MPS MODEL}\label{anmod}

Press \& Schechter (1974) derived, in a rather heuristic manner, the
simple expression
\begin{equation}
N(M,t)\,\der M=\left ({2\over \pi}\right )^{1/2}{\rho_0\over
M^2}{\delc(t)\over \sigma(M)}\left |{\der\ln\sigma\over
\der\ln M}\right | \exp\left [-{\delc^2(t)\over
2\sigma^2(M)}\right ]\,\der M\,
\label{umf}
\end{equation}
for the mass function of haloes in a hierarchical universe endowed
with Gaussian random initial density fluctuations. In equation
(\ref{umf}), $N(M,t)\,\der M$ gives the comoving number density of
haloes with masses in the range $M$ to $M+\der M$ at time $t$,
$\delc(t)$ is the linear extrapolation to the present time $t_0$ of
the critical overdensity for collapse at $t$, $\sigma(M)$ denotes the
r.m.s. mass fluctuation of the linear extrapolation to $t_0$ of the
density field smoothed over spheres of mass $M$, and $\rho_0$ is the
current mean density of the universe.

A more rigorous derivation of the Press--Schechter mass function
(including the correct normalization factor) was provided by Bond et
al. (1991) who also inferred the conditional mass function (see also
Bower 1991). Their theory was subsequently extended
by LC93, who derived the conditional probability that a halo with
$M$ at $t$ ends up within a halo of a larger mass between $M'$ and
$M'+\der M'$ at a later time $t'$. By taking the limit of the latter
conditional probability for $t'$ tending to $t$,\/ LC93 also obtained
the instantaneous halo merger rate,
\begin{eqnarray}
r^{\rm m}_{\rm LC}(M\rightarrow M',t)\,\der M' & = & \left 
({2\over \pi}\right)^{1/2}\,{1\over\sigma^2(M')}\left |
{\der\delc\over \der t}\,\,{\der\sigma(M')\over\der M'}\right|
{1\over[1-\sigma^2(M')/\sigma^2(M)]^{3/2}}\nonumber\\ \label{umr} \\ &
&\times\exp\left\{-{\delc^2(t)\over 2}\left
[{1\over\sigma^2(M')}-{1\over\sigma^2(M)}\right]\right\}\,
\der M'\,,\nonumber
\end{eqnarray}
which provides the fraction of the total number of haloes with $M$
at $t$ which give rise, per unit time, to haloes with masses between $M'$
and $M'+\der M'$ through instantaneous mergers. 

\subsection{Growth rates}\label{hagr}

In the \mps the distinction made between major and minor mergers does
not affect the number density of haloes present at any time. This is
therefore given by the Press--Schechter mass function
(eq.~[\ref{umf}]). However, this distinction substantially modifies the
description of halo growth.

The instantaneous {\it major} merger rate is defined in the same way
as the Lacey--Cole merger rate (eq.~[\ref{umr}]) but is restricted to
mergers above the threshold $\delm$,
\begin{equation}
r^{\rm m}(M\rightarrow M',t)\,\der M'=r^{\rm m}_{\rm LC}
(M\rightarrow M',t)\,\theta[M'-M(1+\delm)]\,\der M'.
\label{mmr}
\end{equation}
with $\theta(x)$ the Heaviside step function.

Note that, due to the divergent abundance of small mass haloes that
can be captured, the Lacey--Cole merger rate $r^m_{\rm LC}(M\to
M',t)\,\der M'$ diverges for $M'-M$ tending to zero (see
eq.~[\ref{umr}]) while the major merger rate defined in equation
(\ref{mmr}) does not have such a divergence.

The integral of this instantaneous major merger rate over the range
of final halo masses gives the rate at which haloes with $M$ at $t$ are
destroyed because of major mergers, i.e., the instantaneous halo
destruction rate,
\begin{equation}
r^{\rm d}(M,t)=\int_M^\infty r^{\rm m}(M\rightarrow M',t)\,\der M'.
\label{desr}
\end{equation}

{}From the major merger rate one can define a useful related quantity,
the instantaneous capture rate, which gives the fraction of the total
number of haloes with $M'$ at $t$ that arise, per unit time, from the
capture with destruction of haloes with smaller masses between $M$ and
$M+\der M$,
\begin{eqnarray}
r^{\rm c}(M'\leftarrow M,t)\,\der M&\equiv& r^{\rm m}(M\rightarrow
M',t)\,{N(M,t) \over N(M',t)}\,\der M\cr \label{capr}
&=& r^{\rm m}_{\rm LC}
(M\rightarrow M',t)\,\theta[M'-M(1+\delm)]\,{N(M,t) \over N(M',t)}\,\der M'
\end{eqnarray}
Note that captured haloes with $M> M'/(1+\delm)$ do not
contribute to this expression because these massive haloes are
not destroyed in the capture and evolve into $M'$ by accretion.

The instantaneous formation rate, i.e., the rate at which haloes with
$M'$ form at $t$ via major mergers, is given by half the integral of
the instantaneous capture rate over the range of captured
haloes more massive than $M'\delm/(1+ \delm)$, 
\begin{equation} 
r^{\rm f}(M',t)= {1\over 2}\,
\int_{M'\delm/(1+ \delm)}^{M'} r^{\rm c}(M'\leftarrow M,t)
\,\der M . \label{forr}
\end{equation}
Indeed, captures with destruction of haloes less massive than
$M'\delm/(1+ \delm)$ do not contribute to the formation of new
haloes. The capture of one single halo of that mass is not enough to
cause the destruction of the capturing halo, while the simultaneous
capture of a large number, giving rise to a substantially more
massive halo than the capturing one and thereby producing its
destruction, is extremely improbable. (The validity of this argument is
confirmed a posteriori.) Therefore, to ensure the formation of new
haloes, captures must involve at least one halo which is more
massive than $M' \delm/(1+ \delm)$. But then haloes with masses
between $\delm M'/(1+ \delm)$ and $M'/2$ will be captured by haloes
with masses essentially in the complementary range from $M'/2$ to
$M'/(1+ \delm)$, and conversely. Therefore, to estimate the halo
formation rate captures must be counted in only one of these
ranges. The two possible estimates corresponding to each range give
essentially the same result in the three cosmologies
analysed. We found no difference in the $n=0$ case
and a constant difference of 1.6 \% and 0.6 \% in the $n=-2$ and
$n=-1$ cases, respectively. A slightly better estimate, adopted in
expression (\ref{forr}), is still provided by the arithmetic mean of
both estimates.

\begin{figure}
\psfig{file=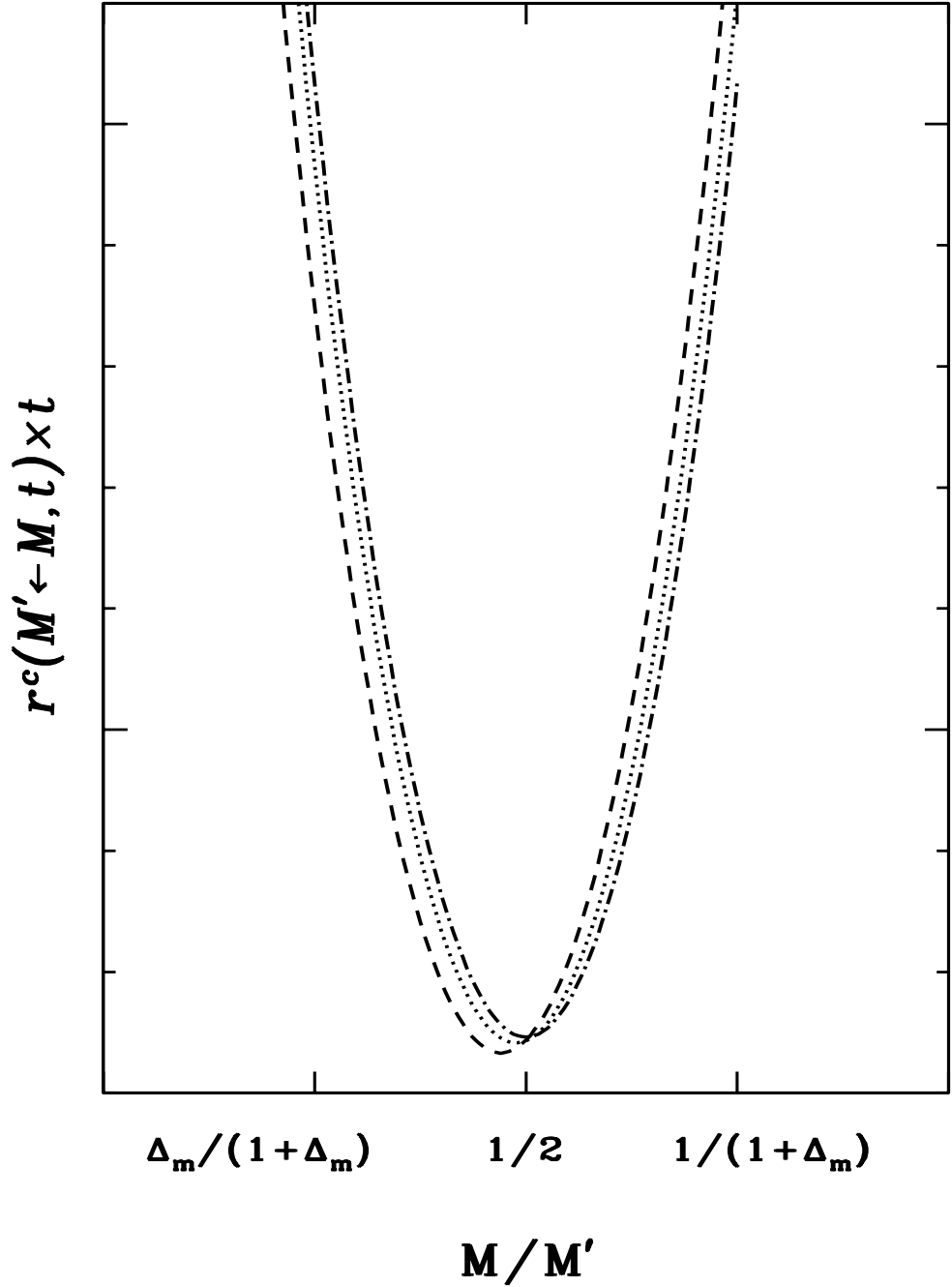,height=21cm,width=\textwidth}
\caption{Shape around $M'/2$ of the instantaneous capture rate, in
arbitrary units, for $M'=M_\ast$. The captured mass $M$ on the $x$-axis is
normalized to $M'$ so to have time-independent results. The 
curves correspond to the three cosmologies analysed ($n=-2$
in dashed line, $n=-1$ in dotted line, and $n=0$ in dot-dashed
line). A similar behaviour is found for any other
value of $M'$.}
\label{fcapr}
\end{figure}

The validity of binary approximation used in the previous
reasoning is shown not only by the similarity between the two
alternative estimates of the formation rate, but mainly by the shape
of the capture rate in the range $M'\delm/(1+ \delm)< M<
M'/(1+\delm)$: this function is very nearly symmetric around $M'/2$
(see Fig.~\ref{fcapr}). It must be emphasized that in the present
theory the formation of a new halo is an instantaneous event that
corresponds to the capture by some predecessor of a similar massive
object. This massive capture is clearly less probable than the capture
of a less massive object. It is therefore understood that halo
formation corresponds to rare binary mergers between similar massive
haloes while accretion corresponds to much more frequent, often
multiple, minor mergers.

In the \mpsc the total mass increase rate of haloes with $M$ at $t$
splits into two contributions, one due to major mergers and the other
to accretion. Here, we are particularly concerned with the
latter. The instantaneous mass accretion rate, i.e., the rate at which
haloes with $M$ at $t$ increase their masses due to minor mergers, is
\begin{equation}
r^{\rm a}_{\rm mass}(M,t)=\int_M^{M(1+\delm)}\,(M'-M)
\,r^{\rm m}_{\rm LC}(M\rightarrow M',t)\,\der M'\, . \label{accr}
\end{equation}

This rate, as with all preceding ones, is intended to globally
characterize haloes with $M$ at $t$. It measures the {\it
expectation value}, for any halo, of the instantaneous mass
increase rate due to minor mergers, or equivalently, by the ergodicity
condition, the {\it average} of the real instantaneous mass increase
rate due to minor mergers over all these haloes. Nevertheless, since
accretion is a very common event involving a number of relatively
small mass haloes, equation (\ref{accr}) should also provide a
reasonable approximation for the true instantaneous mass increase rate
experienced by any halo with $M$ at $t$ due to
accretion. Consequently, the solution of the differential equation
\begin{equation}
{\der M(t)\over\der t}=r^{\rm a}_{\rm mass}[M(t),t], \label{act}
\end{equation} 
with $M(t_{\rm i})=M_{\rm i}$, giving the {\it mean} $M(t)$ trajectory
followed between two major mergers by haloes with $M_{\rm i}$ at
$t_{\rm i}$, should also be a reasonable approximation for the true
$M(t)$ track followed by any halo. For this reason, we shall call this
``the theoretical accretion track''. The validity of this
approximation will be checked in Section \ref{mactr}.

It can be argued that in the \mps the parameter $\delm$ plays the same
role as the mass resolution used in the EPS model when building Monte
Carlo merger trees and that, in the present model, the formation of
new haloes through binary major mergers between similarly massive
haloes plays the same role as merger nodes in the Monte Carlo
simulations. In both cases only captures above some mass cut-off are
seen as true mergers while those below it are regarded as contributing
to accretion. However, the potential of the \mps to solve the problems
mentioned in Section \ref{intro} does not rely on this distinction
alone, but also on the new definitions of halo destruction and
formation that accompany it. In the \mps there are haloes which are
destroyed or formed at the exact moment in which they are
analysed. Therefore, it is possible to define and derive the
instantaneous destruction and formation rates (eqs.~[\ref{desr}] and
[\ref{forr}]) with no counterpart in the EPS model and, from them (see
next subsections), the PDFs of formation and destruction times and of
progenitor masses that solve problems (ii) and (iii). Furthermore,
because halo formation takes place at definite times in which the
progenitors and the newly formed haloes coincide, we can identify such
events as {\it true\/} merger nodes. Finally, the mass increase of a
halo due to the cumulative effect of minor mergers that occur at any
time after its formation is consistently taken into account in the
\mps through the instantaneous mass accretion rate (eq.~[\ref{accr}])
or the theoretical accretion track derived from it (eq.~[\ref{act}])
which also solves problem (i).

\subsection{Formation times}\label{hftpm}

To derive the PDF of halo formation times we shall assume that the
mass evolution $M(t)$ of a halo since the last major merger that gave
rise to it is well described by the theoretical accretion track
(eq.~[\ref{act}]). In this case, the cumulative number density of
haloes at $t_{\rm i}$ with masses in the arbitrarily small range
$M_{\rm i}$ to $M_{\rm i}+\delta M_{\rm i}$ which pre-exist at an
early time $t<t_{\rm i}$, i.e., the spatial number density of haloes
which evolve by accretion from $t$ to $t_{\rm i}$ ending up with a
mass between $M_{\rm i}$ and $M_{\rm i}+\delta M_{\rm i}$ is (Manrique
\& Salvador-Sol\'e 1996; Manrique et al.~1998)
\begin{equation}
N_{\rm pre}(t)= N[M_{\rm i},t_{\rm i}]\,\delta
M(t)\,\exp\left\{-\int_t^{t_{\rm i}} r^{\rm f}[M(t'),t')]\, \der
t'\right\},\label{pre}
\end{equation}
where
\begin{equation}
\delta M(t)= \delta M_{\rm i}\,\exp\left[-\int_t^{t_{\rm i}}{\partial 
r^{\rm a}_{\rm mass}(M,t')\over\partial M}\biggr|_{M=M(t')}\,\der t'
\right]\label{mel}
\end{equation}
is the mass element at $t$ that evolves into $\delta M_{\rm i}$ at
$t_{\rm i}$ following theoretical accretion tracks. Therefore, the PDF
of formation times for haloes with masses between $M_{\rm i}$ and
$M_{\rm i}+\delta M_{\rm i}$ at $t_{\rm i}$ is given by
\begin{equation}
\Phi_{\rm f}(M_{\rm i},t)\equiv{1\over N_{\rm pre}(t_{\rm i})}\,{\der
N_{\rm pre} \over\der t}=r^{\rm f}[M(t),t]\,
\exp\biggl\{-\int_t^{t_{\rm i}} r^{\rm f}[M(t'),t']\,\der
t'\biggr\}\,,
\label{dft}
\end{equation}
where $N_{\rm pre}(t_{\rm i})=N(M_{\rm i},t_{\rm i})$. The median for
this distribution defines the typical formation time $t_{\rm f}$ of
haloes with $M_{\rm i}$ at $t_{\rm i}$. This follows from the usual
definition that the typical formation time of a population is the
epoch in which its abundance was a factor $e$ smaller. Similarly (see
Manrique \& Salvador-Sol\'e 1996 and Manrique et al.~1998), one can
derive the PDF of halo destruction times and the associated typical
destruction time $t_{\rm d}$. Finally, the typical survival time of
haloes with $M_{\rm i}$ at $t_{\rm i}$ is taken as the difference
$t_{\rm d}-t_{\rm i}$.

In estimating the mass evolution of haloes between major mergers by
means of their theoretical accretion tracks we neglect the diffusion
of true accretion tracks (in the mass vs.~time diagram) caused by
random minor mergers. Were this diffusion negligible or exactly
symmetric with regard to the mean $M(t)$ trajectory given by the
theoretical accretion track, the following conservation equation of
the number density of haloes per unit mass along accretion tracks
would be satisfied
\begin{equation}
{\der\ln N(M,t)\over \der t}+{\partial r^{\rm a}_{\rm mass}(M',t)\over
\partial M'}\biggl|_{M'=M}-r^{\rm f}(M,t)+r^{\rm d}(M,t)=0\,.
\label{coec}
\end{equation}
This conservation equation was used by SSM (suggested by Manrique et
al.~1998) to derive the formation rate from quantities $N(M,t)$,
$r^{\rm a}_{\rm mass}(M',t)$, and $r^{\rm d}(M,t)$. Note that this
estimate of the formation rate alternative to equation (\ref{forr}) is
independent of the binary major merger approximation but may be
affected, in turn, by any asymmetric diffusion of accretion
tracks. Since binary approximation is very accurate, the
comparison between these two alternative estimates of the formation
rate provides quantitative information about the true asymmetry of
diffusion.

\begin{figure}
\psfig{file=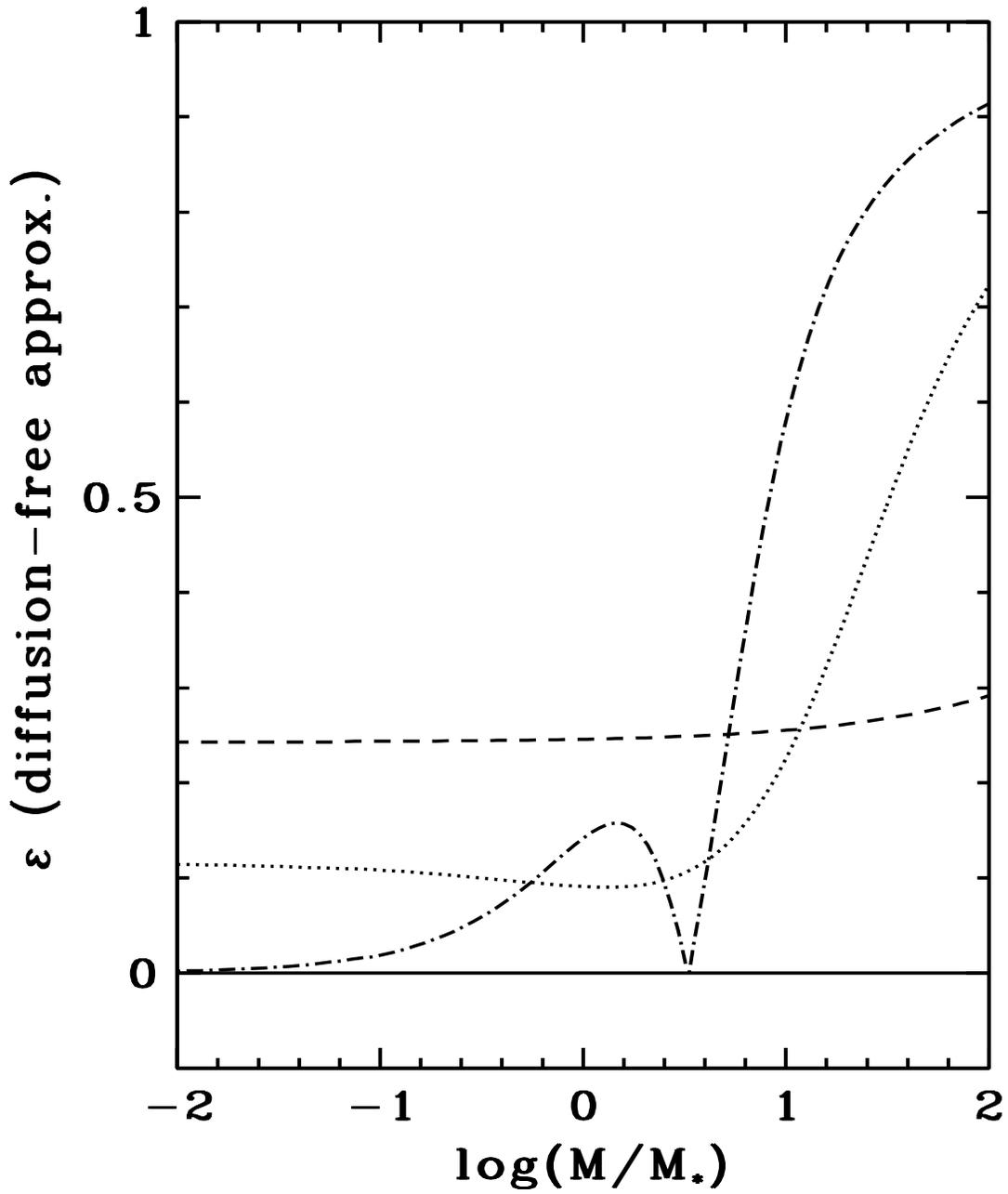,height=21cm,width=\textwidth}
\caption{Absolute value of the relative difference, as a function of
halo mass, between the formation rates estimated from condition
(\protect\ref{coec}) and equation (\protect\ref{forr}) for the three scale-free
cosmologies analysed (same line coding as in Fig.~\protect\ref{fcapr}). Halo
masses on the $x$-axis are normalized to the characteristic mass
$M_\ast$ defined in equation (\protect\ref{mstar}) so as to have time-independent
results.}
\label{fgcon}
\end{figure}

In Figure~\ref{fgcon} we show the absolute value of the relative
difference between the two estimates of the formation rate
(eqs.~[\ref{forr}] and [\ref{coec}]) as a function of halo mass for
several cosmologies. For large $n$ there is good
agreement at small masses but there is a significant deviation at
large masses, while for small $n$ the difference is rather insensitive
to mass, although it is significant at all values. We conclude that the
effects of diffusion cannot be neglected and,
consequently, that the only reliable estimate of halo formation rate
is given by equation (\ref{forr}). The effects of diffusion on the
distribution of formation times (eq.~[\ref{dft}]) is 
examined in Section \ref{resul}.

\subsection{Progenitor masses}\label{pm}

Finally, the PDF of progenitor masses can be derived by taking
into account that major mergers are essentially binary. Given a halo
formed at $t$ with mass $M$, the probability that the most massive or
primary progenitor has a mass in the range $M_1$ to $M_1+\der M_1$ at
$t$ is given by
\begin{eqnarray}
\Phi_{\rm p}(M_1,M)\,\der M_1&\equiv& {r^{\rm c}(M\leftarrow M_1,t)\over
r^{\rm f}(M,t) }\,\der M_1\cr &=&2\,G(M_1,M)\,\der M_1\left[ 
\int_{M\delm/(1+ \delm)}^{M/(1+\delm)}
G(\widetilde M,M)\,\der \widetilde M\right]^{-1}
\label{mpro}
\end{eqnarray}
with
\begin{equation}
G(M',M)={1\over M'\,\sigma^2(M')}\left |{\der\sigma(M')\over\der M'}
\right|\left[1-{\sigma^2(M)\over\sigma^2(M')}\right]^{-3/2} \, .
\label{fung}
\end{equation}
In equation (\ref{mpro}), $M_1$ takes values in the range between
$M/2$ and $M/(1+ \delm)$. The PDF for the mass $M_2$ of the secondary
progenitor is given, in turn, by the same expression (\ref{mpro}) but
replacing $M_1$ by $M_2$ which takes values in the range $\delm M/(1+
\delm)$ to $M/2$. Since the capture rate is essentially symmetric
around $M/2$ (see Fig.~\ref{fcapr}) the typical masses of the two
progenitors can be computed by taking the median value $M_1$ for the
distribution (\ref{mpro}) of the primary progenitor and then the
difference $M_2=M'-M_1$ for the secondary one, or conversely. Notice
that the PDFs of primary and secondary progenitor masses in the MPS
theory are independent of time (see eqs.~[\ref{mpro}] and
[\ref{fung}]).

The number and total mass of progenitors of a halo are fully
determined in the present theory thanks to binary
approximation. This approximation is amply justified by
the shape of the instantaneous capture rate. Nonetheless, all
predictions based on it will be carefully checked in Section
\ref{resul}. The same approximation is often used when
constructing Monte Carlo merger trees from the EPS model. There is,
however, an important difference between the two cases. In the \mpsc
major mergers yielding the formation of new haloes involve only
similar massive progenitors and occur effectively at the
formation time, while in the EPS model, mergers involve resolved
progenitors of any mass ratio and can occur at any moment between
two consecutive nodes of the merger tree. Therefore the validity of
binary approximation is not so obvious in the latter case. Of
course, one may enforce it by taking a small enough time step and a
sufficiently large mass resolution, but this tends to reproduce
the optimal conditions encountered, by construction, in the \mpsd

\section{N-BODY SIMULATIONS}\label{nbsim}

\subsection{The data}

To check that the behaviour of the \mps is correct, we have used the
outcome of the cosmological $N$-body simulations performed by LC94 and
used by these authors to test the EPS model. Here we only provide a
brief summary of the most relevant aspects of these simulations
concerning the present work (see LC94 for a detailed description).

Simulated data correspond to three Einstein--de Sitter (i.e.,
$\Omega=1$, $\Lambda=0$) universes with a power-spectrum of initial
density fluctuations given by a power law, $P(k) \propto k^n$, with
spectral index $n$ equal to $-2$, $-1$, and $0$. The simulations were
performed using the $P^3M$ code of Efstathiou et al.~(1985) with
$128^3$ particles and a mesh of $256^3$ points. The output times were
selected so that the characteristic mass $M_\ast(t)$ increases by a
factor of $\sqrt2$ between each pair of successive output times, where
$M_\ast(t)$ is defined by the relation
\begin{equation}
\sigma[M_*(t)]=\delc(t)=\delc(t_0)\,{a(t_0)\over a(t)}, \label{mstar} 
\end{equation}
with $a(t)$ the expansion factor of the universe.

LC94 found good agreement between the analytical predictions of the
EPS model, for $\delc(t_0)=1.69$ and a top-hat filter (with smoothing
scale $R$ related with the filter mass through
$M=\displaystyle{4\pi\over 3}\rho_0 R^3$), and the results of their
simulations when haloes were selected by means of the conventional
friends-of-friends algorithm with a dimensionless linking length of
$b=0.2$. In the present work, we use the same strategy to identify
haloes from simulations. The same filtering window and the same value
of the critical overdensity are also used to derive the theoretical
predictions from the \mpsd LC94 found better
fits to the {\it mass function\/} of simulated haloes when
$\delc(t_0)$ was allowed to vary. In any event, the best-fit was
obtained for a value which was still very close to the standard of
$1.69$. (Using this standard value LC94 found that the theoretical
mass function overestimates the abundance of haloes in \nbody
simulations, particularly at the lower and higher mass ends, by a
factor never exceeding $2$; see Fig.~1 of LC94.)

In the three simulated scale-free universes the evolution of 
structure is {\it self-similar}. In this case one has
\begin{equation}
\sigma(M) \propto M^{-(n+3)/6},
\end{equation}
implying that $a(t)$ changes by a factor $2^{(n+3)/12}$ between each
successive output time of the simulations. Consequently, the output
time step for each individual simulation is constant in logarithmic
scale, although it is longer in simulations with a larger value of
$n$. As pointed out by LC94, the advantage of having self-similar
universes is that objects identified at different times $t$ but with
the same value of the scaled mass $M(t)/M_\ast(t)$ are
indistinguishable. Therefore we can normalize all masses $M$ to the
characteristic mass $M_\ast$ at the corresponding time $t$ and combine
the results obtained from bins of identical normalized mass $\widetilde
M$ for several output times, thereby drastically reducing the statistical
noise in the empirical quantities derived from simulations.

\subsection{Tracing halo evolution}\label{trhev}

Here we describe the general procedure that we followed to trace the
past and future evolution of haloes present at a given output time in
the simulations. The first practical problem encountered concerns the
stability of haloes in simulations. As it is well known, halo-finding
algorithms, such as the friends-of-friends technique, have several
intrinsic shortcomings which may distort the evolution of simulated
haloes: those particles that are close, although gravitationally
unbound, to a halo may be assigned to it. Conversely, gravitationally
bound particles may be excluded when their orbits have led them far
from the identified halo region. To minimize these effects we followed
LC94 and considered only haloes with at least $20$ particles. Another
difficulty is that the outcome of simulations is only saved at a
discrete set of times $t_k$ while the MPS theory deals with
instantaneous major mergers that occur at definite times which do not
coincide in general with any output time $t_k$. Since the most
important quantities in the EPS theory refer to arbitrary time
intervals this difficulty could be avoided in the LC94 study. Note,
however, that this difference between the two studies does not
reflect any shortcoming of the \mps over the EPS theory, but
simply that the former provides a finer description of the dark-matter
aggregation process.

We must assess whether a halo with mass $M$ at an output time $t$
formed in a major merger that took place {\it after the previous
output time} $t'$ or has evolved by accretion (i.e., it has only
experienced minor mergers) since then. Suppose we identify the halo
$M_1$ at $t'$ harbouring the largest number of particles which end up
within $M$ at $t$. We will call it the most massive predecessor at
$t'$ of the halo $M$ at $t$ (not to confuse this with the primary
progenitor). The theoretical threshold $\delm$ applies in the case of
$t$ and $t'$ infinitely close, while the masses $M$ and $M_1$ refer to
two distinct times. So even when $(M-M_1)/M$ is larger than $\delm$
the evolution may have taken place by accretion. This is true when the
remaining mass $M-M_1$ is distributed among many small lumps. On the
other hand, even in the case of binary major mergers, the particles
located in the final object $M$ will not be found within only two
haloes at $t'$: apart from the major merger, there is some accretion
onto the two progenitors before the merger and onto the final halo
after it. In other words, neither of these two simple criteria (i.e.,
the condition $(M-M_1)/M>\delm$ and the location of the particles
belonging to halo $M$ in only two predecessors) can identify halo
formation. For this reason, we applied an improved combination of
both. Concretely, we took into account that major mergers yielding the
formation of new haloes are binary events among coeval haloes that
have increased their masses by accretion since $t'$ up to the merger
time in some amounts that are, in a first approximation, proportional
to the respective masses at $t'$. In practice this is implemented as
follows. We find the first and second most massive predecessors $M_1$
and $M_2$ at $t'$ of the halo $M$ at $t$. We then compare the ratio
$N_2/N_1$ with the threshold $\delm$, where $N_1$ and $N_2$ are the
number of particles in haloes $M_1$ and $M_2$, respectively, with
which these two predecessors contribute to the final halo $M$. When
$N_2/N_1>\delm$, we consider that $M$ formed in a major merger between
$t'$ and $t$ and identify the two predecessors $M_1$ and $M_2$ as the
progenitors which were destroyed in the event. On the contrary, when
$N_2/N_1\le\delm$ we consider that $M$ is the result of the evolution
by accretion of the most massive predecessor $M_1$.

We must also assess whether a halo with mass $M'$ at an output time
$t'$ was destroyed in a major merger {\it before the next output
time} $t$ or, alternatively, has evolved by accretion from $t'$ to
$t$. The procedure followed in this case is similar to that
explained above. We find the set of haloes where the particles in $M'$
end up at the later time $t$ and select from this set the halo $M$
that contains the greatest number of particles originally in
$M'$. This halo is identified as the successor of $M'$. (When the
successor of $M'$ had less than $50\%$ of its particles, then $M'$ was
discarded from our sample of putative haloes since it was not stable
enough.)  After identifying the successor $M$ at $t$ of $M'$ at
$t'$, we find the most massive predecessors of $M$ at $t'$. There
are then two possibilities: 1) $M'$ is not the most massive predecessor of
$M$, in which case $M'$ has certainly been captured by a more massive
halo between $t'$ and $t$, being destroyed in the event; and 2) $M'$
is the most massive predecessor of $M$, in which case we must identify
the second most massive predecessor of $M$ to assess, as
explained above, whether $M$ was formed (and $M'$ destroyed)
between $t'$ and $t$ or $M'$ has evolved into $M$ by accretion.

Following these prescriptions we can readily compute the fraction of
haloes within a given mass bin which were formed or destroyed at
some moment between two consecutive output times of the simulation. We
can also identify haloes evolving by accretion and compute their mass
evolution along a series of discrete times. Finally, we can identify
the two progenitors of a halo and estimate its formation (or
destruction) time.

\section{RESULTS}\label{resul}

\subsection{Formation, destruction, and mass accretion rates}\label{grat}

To evaluate the empirical formation rate we compute the fraction of
haloes in a mass bin formed between two consecutive output times
$t_{k-1}$ and $t_k$ and divide it by $\log(t_k/t_{k-1})$. This gives
an estimate of the product $r^{\rm f}(\widetilde{M},t_k)t_{k}$,
where $\widetilde{M}$ is the central value of the normalized mass bin
at $t_k$. Similarly, the destruction rate is evaluated by computing
the fraction of haloes in a mass bin at $t_k$ which are destroyed in a
major merger between the consecutive output times $t_k$ and $t_{k+1}$
divided by $\log(t_{k+1}/t_{k})$ which gives an estimate of the
product $r^{\rm d}(\widetilde{M},t_k)t_{k}$. In both cases, we
average the results obtained from all pairs of consecutive output
times by weighting their contribution in proportion to the number of
haloes found in the corresponding normalized mass bin. The statistical
error of this combined value is estimated as the weighted averaged
Poisson errors of the values corresponding to each pair of consecutive
output times.

To estimate the mass accretion rate, we first determine the set of
haloes in a given mass bin at $t_k$ that are not destroyed (i.e.,
which evolve by accretion) between $t_k$ and $t_{k+1}$. Then, the
average mass increase from $t_k$ to $t_{k+1}$ of haloes in such a set
normalized to $M_\ast(t_k)$ and divided by $\log(t_{k+1}/t_k)$ yields
an estimate of $[\der M/\der \log(t)]/M_\ast$, or equivalently, an
estimate of the dimensionless quantity $r^{\rm a}_{\rm
mass}(M,t_k)t_{k}/M_\ast(t_{k}$), as we wanted. Finally, we
average the results obtained for different times $t_k$ in the same way
as for the formation and destruction rates. For the error associated
with this average, we take the weighted averaged r.m.s. mass increment
of haloes in the mass bin divided by the square root of the total
number of haloes in the bin.

\begin{figure}
\psfig{file=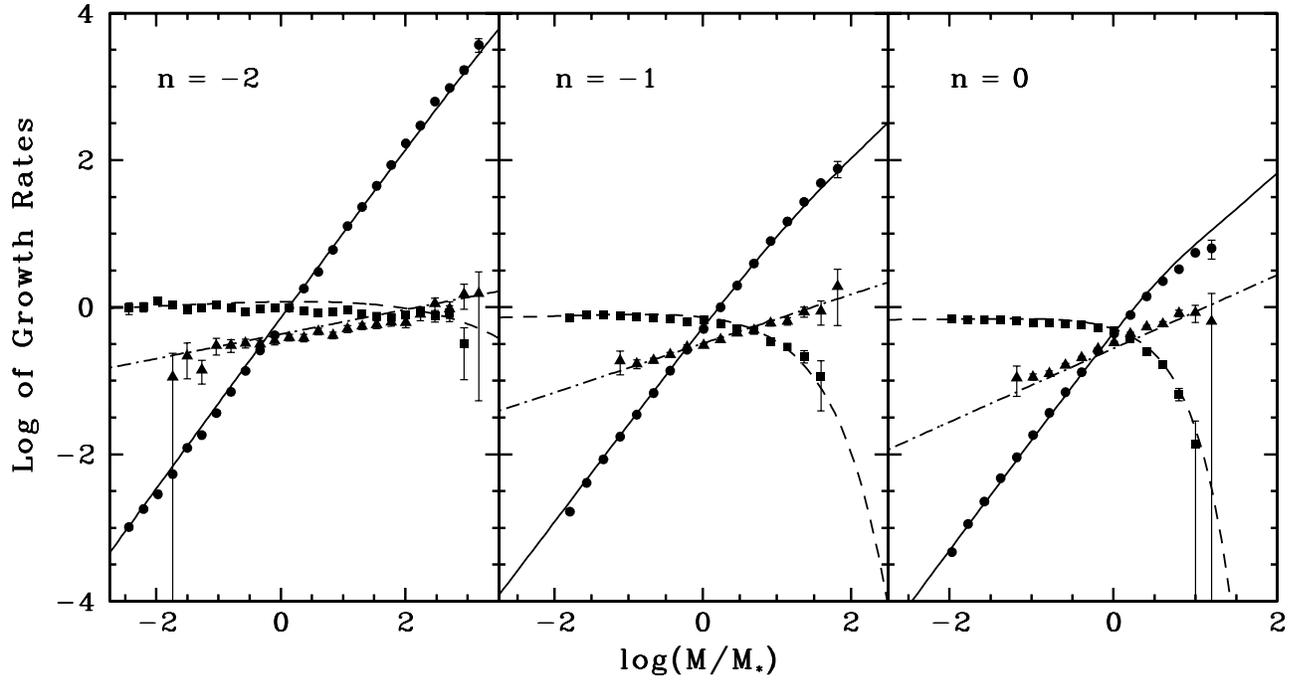,height=14cm,width=18.5cm,angle=-90}
\caption{Comparison between the growth rates predicted by the \mps
(lines) and drawn from the simulations (symbols) using a logarithmic
time-increment equal to the output time step of the corresponding
simulation for the three scale-free cosmologies analysed: $n=-2$ (left
panel), $n=-1$ (central panel), and $n=0$ (right panel). Theoretical
destruction rates are in dashed lines (squares for the empirical
results), formation rates in dot-dashed lines (triangles), and mass
accretion rates in solid lines (circles). Error bars and symbols have
similar sizes except at both mass ends.}
\label{rates1}
\end{figure}

\begin{figure}
\psfig{file=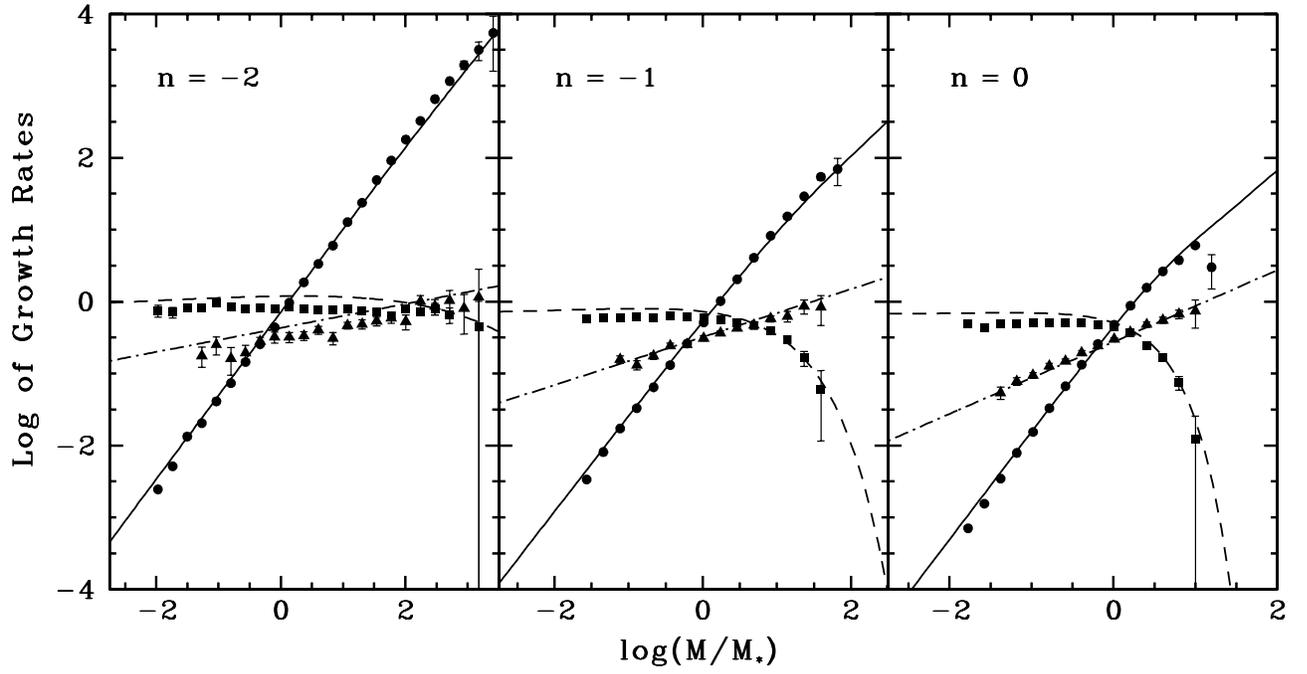,height=14cm,width=18.5cm,angle=-90}
\caption{Same as Figure \protect\ref{rates1} for a logarithmic time-increment
equal to two output time steps of each simulation.}
\label{rates2}
\end{figure}

Note that, in practice, the empirical time-derivatives that give the
different growth rates are estimated by means of the finite increment
approximation. Unfortunately, the size of the logarithmic
time-increment used cannot be chosen arbitrarily small since the
minimum available value is fixed by the output time step of the
simulation. To check the finite increment approximation we present the
results obtained using two different logarithmic time-increments: one
equal to the output time step of the simulation (Fig.~\ref{rates1})
and the other equal to two output time steps (Fig.~\ref{rates2}). As
can be seen, there is an excellent agreement between the theoretical
and empirical growth rates for the whole range of halo masses that can
be examined (almost four decades).  The very slight deviations that
can be detected (as in the case of the destruction rate) increase with
increasing time-increments, which strongly suggests that they are
caused by the finite-increment approximation.

The impressive agreement between theory and simulations is so far not
surprising since the growth rates analysed were computed, after
distinguishing between major and minor mergers, from the Lacey--Cole
merger rate (LC93) which itself is in excellent agreement with
$N$-body simulations (LC94). These results confirm, however, the
overall consistency of the MPS prescriptions and the correctness of
the practical procedure used to identify the formation and destruction
of haloes in the simulations.

\subsection{Accretion tracks}\label{mactr}

When comparing simulated and theoretical accretion tracks we are
concerned with two main aspects: first, whether the real accretion
tracks of individual haloes arising from random minor mergers
substantially deviate from the theoretical track, in other words,
whether the diffusion of real accretion tracks is important, and
second, whether the theoretical accretion track is a good
estimate of the {\it average} mass evolution followed by haloes
between two consecutive major mergers, i.e., whether the diffusion is
symmetric with regard to the theoretical mean track.

\begin{figure}
\psfig{file=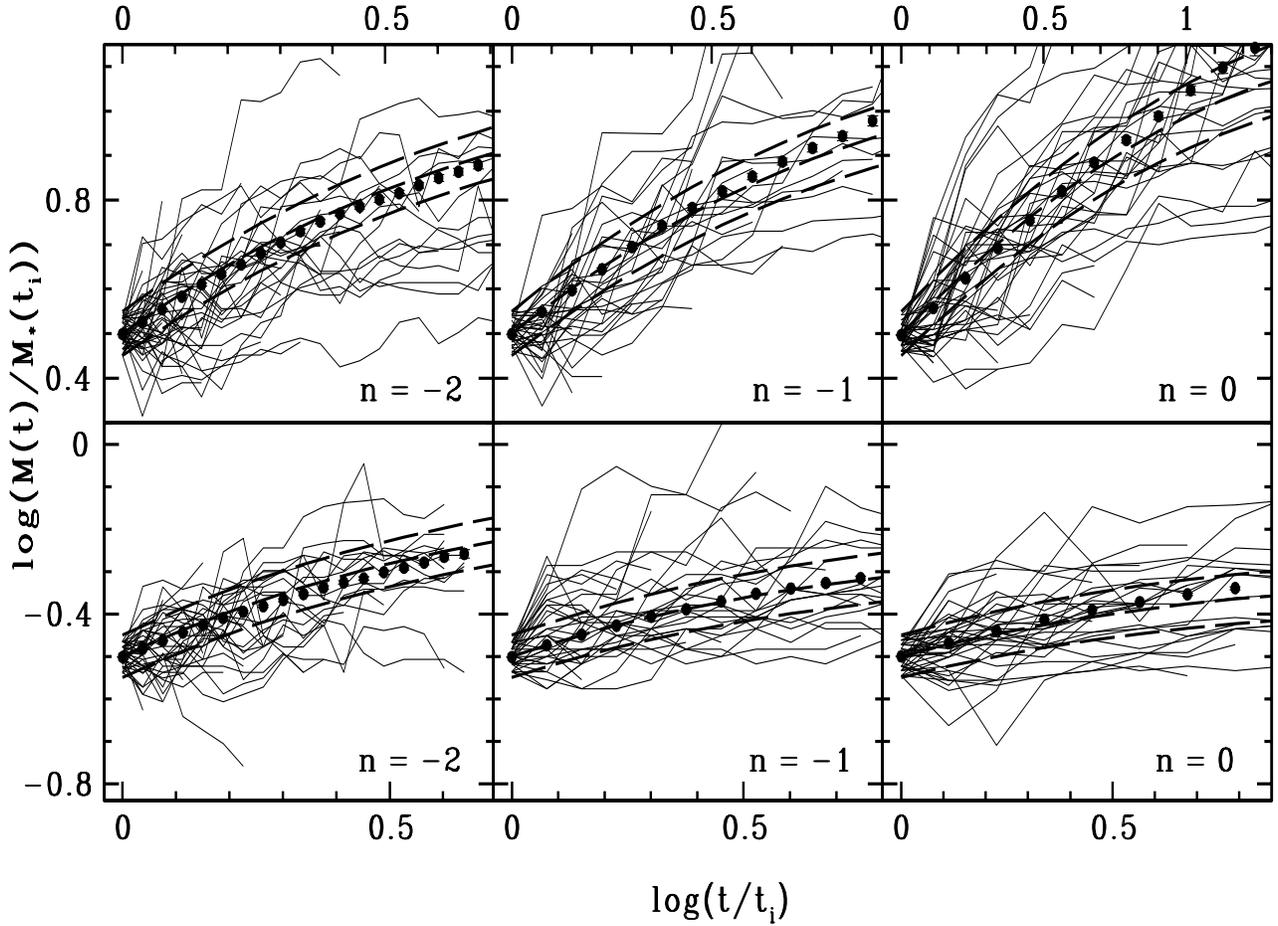,height=18cm,width=18.5cm,angle=-90}
\caption{Accretion tracks followed by randomly selected haloes with
initial masses at $t_{\rm i}$ in logarithmic bins of $0.1$ width
centred on $M=0.3M_\ast$ (lower panels) and $M=3.0M_\ast$ (upper
panels) in the three cosmologies analysed. In thin lines tracks
drawn from simulations.  In thick dashed lines the theoretical
accretion tracks for the central mass as well as for the two masses
bracketing each initial mass bin. Dots show the time evolution of the
average mass of simulated haloes initially in each mass bin. Error
bars are much smaller than the dot size.}
\label{figtra}
\end{figure}

For each individual halo with $M_{\rm i}$ at $t_{\rm i}$ which does
not satisfy the condition of an imminent major merger, we identify its
successor at the next output time, and repeat the same procedure
starting with this successor. We iterate the process until a successor
is destroyed in a major merger. In this way we can predict the evolution
due to accretion of the quantity $M(t)/M_\ast(t_{\rm i})$
that corresponds to any particular halo by taking the masses $M(t)$ equal
to the masses of the successors found along the series of respective
output times $t_k> t_{\rm i}$.  Figure~\ref{figtra} shows, for the
three universes analysed, the mass evolution due to accretion of
randomly selected haloes with initial normalized masses at $t_{\rm i}$
(corresponding to different output times of the simulation) in the
logarithmic bins around $\widetilde M=0.3$ and $\widetilde M=3.0$. For
comparison, we plot the theoretical accretion tracks corresponding to
the central and the two limiting masses in each initial bin. As can be
seen, the tracks of the simulated haloes show the same trend as those
predicted by the theory. The former are, of course, affected by small random
deviations (some of them quite notable because of minor mergers close
to $\delm$), but in general the diffusion is rather moderate, as
expected, because of the high frequency of minor mergers (particularly of
very small mass) acting continuously between consecutive output
times. This is apparent from the slow increase with time of the
scatter of true tracks around the mean. At the ending time in each
plot corresponding to two typical survival times of haloes in the initial bin,
more than $20$--$30\%$ of the surviving simulated tracks still remain
inside the theoretical tracks bracketing each bin. (These percentages
correspond to the global samples, not the random subsamples plotted in
Fig.~\ref{figtra}).

{}From Figure~\ref{figtra} one can also see that the diffusion is
slightly asymmetric. The departure of the empirical average track
relative to that of the theoretical mean is towards small masses when
$n=-2$, towards large masses when $n=0$, and intermediate between the
two when $n=-1$. None the less, these deviations are quite
moderate at least until two survival times, the only exception being
the most massive bin of the $n=0$ case. Note that such a deviation
begins to be marked when the evolving normalized mass approaches
unity. This is in overall agreement with our previous results on the
evidence of asymmetric diffusion drawn from Figure~\ref{fgcon}.

\subsection{Distribution of formation times}\label{disft}

To determine the formation time of a simulated halo we follow its
accretion-driven evolution back in time until a major merger is
reached. In practice this is done in the following way. For a halo of
mass $M$ at $t_{\rm i}$ equal to some output time $t_k$ we identify
its two most massive progenitors at $t_{k-1}$ and check (as explained
in Section \ref{nbsim}) whether a major merger took place between
$t_{k-1}$ and $t_k$. When this is not the case, we repeat the
procedure for the most massive predecessor of $M$ at $t_{k-1}$. By
iterating this process we can follow the evolution of the successive
most massive predecessors until the condition for a major merger is
fulfilled. We then take the geometric mean between the two consecutive
output times bracketing the major merger event as the formation time
of the original halo with $M$ at $t_{\rm i}$. In this way we can
obtain the formation time of all haloes in a fixed mass bin at $t_{\rm
i}$. The total fraction of haloes formed at distinct time intervals
divided by the logarithmic output time step of the simulation yields a
direct estimate of the product $\Phi_{\rm f}(M,t) t$. The
results obtained for distinct values of $t_{\rm i}$ (corresponding
output times of the simulation) are finally averaged by weighting each
individual contribution according to the total number of haloes from
which the distribution is calculated (only those reliable
distributions drawn from at least $50$ objects are considered). The
uncertainty associated to each bin of formation times is given by the
weighted averaged Poisson errors from individual distributions.

\begin{figure}
\psfig{file=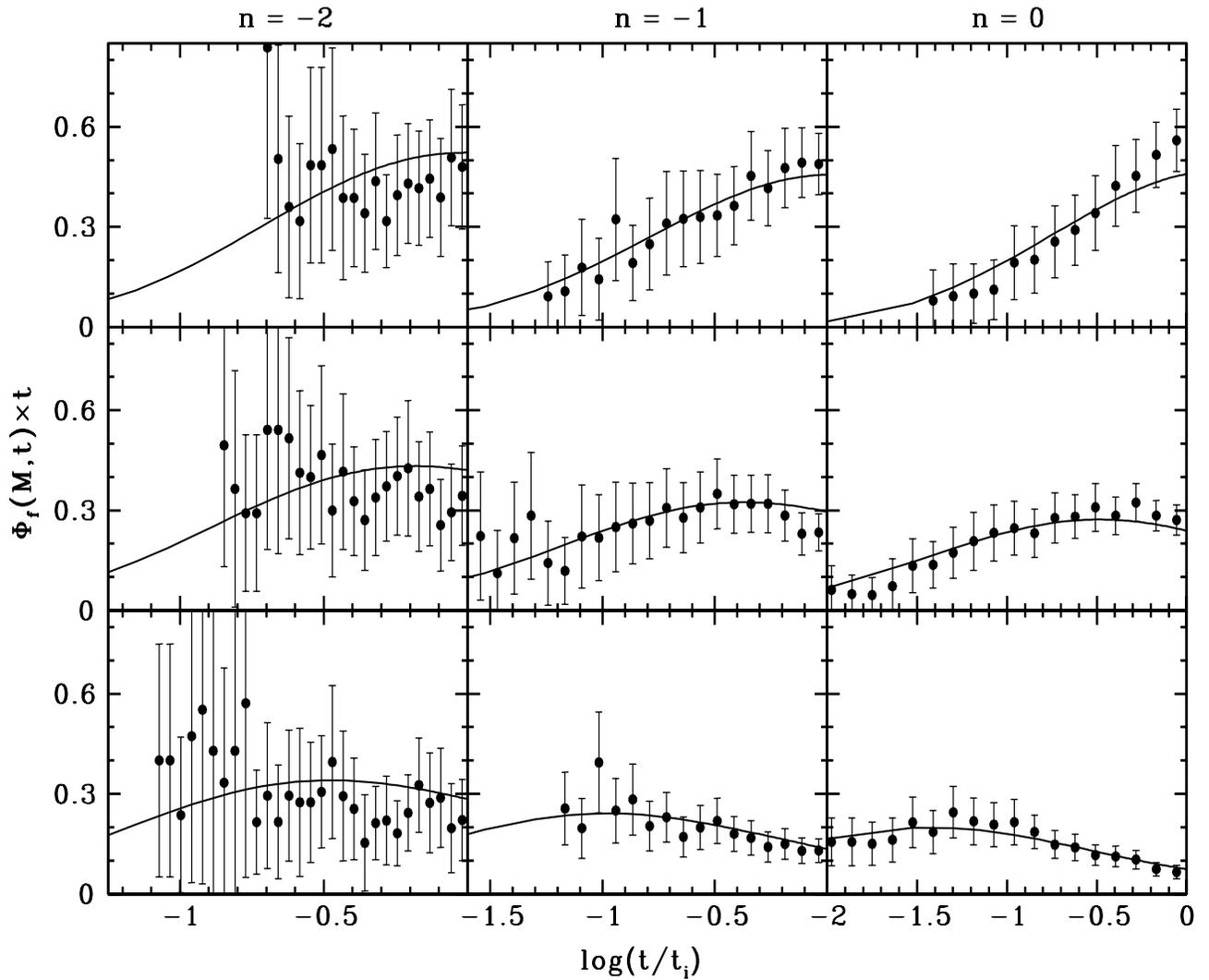,height=16cm,width=18.5cm,angle=-90}
\caption{Comparison between theoretical (solid lines) and empirical
(dots) formation time PDFs for haloes of three distinct masses
(2.8 $M_\ast$ in the top panels, 0.7 $M_\ast$ in the middle panels, and
0.07 $M_\ast$ in the bottom panels) in the three cosmologies analysed.}
\label{timedi}
\end{figure}

In Figure~\ref{timedi} we show the empirical formation time PDFs of
haloes with three different masses at $t_{\rm i}$ compared with the
MPS predictions for the three universes analysed. As can be seen,
there is also very good agreement for the full ranges of time that can
be examined from the simulations in each particular case. We remind
that the MPS prediction assumes haloes following the theoretical mean
accretion track. We therefore conclude that the small diffusion of
accretion tracks (shown in Figure~\ref{figtra}) has a negligible effect
on these PDFs. This conclusion holds even for large masses and very
small formation times (of about one hundredth of the reference time
$t_{\rm i}$) as reached in the case of $n=0$, when the effects of
diffusion are the most marked.

\subsection{Distribution of progenitor masses}\label{dhpm}

For any newly formed halo (i.e., satisfying the condition of a recent
major merger) we can store the masses of its two most massive
predecessors which, according to our definition in Section
\ref{trhev}, are considered its two progenitors. In this way the
empirical PDF of progenitor masses can be computed for haloes selected
in a given normalized mass bin at various output times. These
results are then averaged following the same procedure as described in
the previous subsection to derive the empirical PDF of formation
times. Error bars are also calculated in the same manner.

According to our practical identification of halo formation explained
in Section \ref{trhev}, the mass ratio between the two most massive
progenitors at formation is taken to be equal to the ratio $N_1/N_2$
between their respective number of particles at the previous
output time that are found within the newly formed halo at the next
output time. Likewise, the mass ratio at formation between any of
these progenitors and the new halo is taken to be equal to the ratio $N_{\rm
p}/(N_1+N_2)$ where $N_{\rm p}$ stands for $N_1$ or $N_2$.

\begin{figure}
\psfig{file=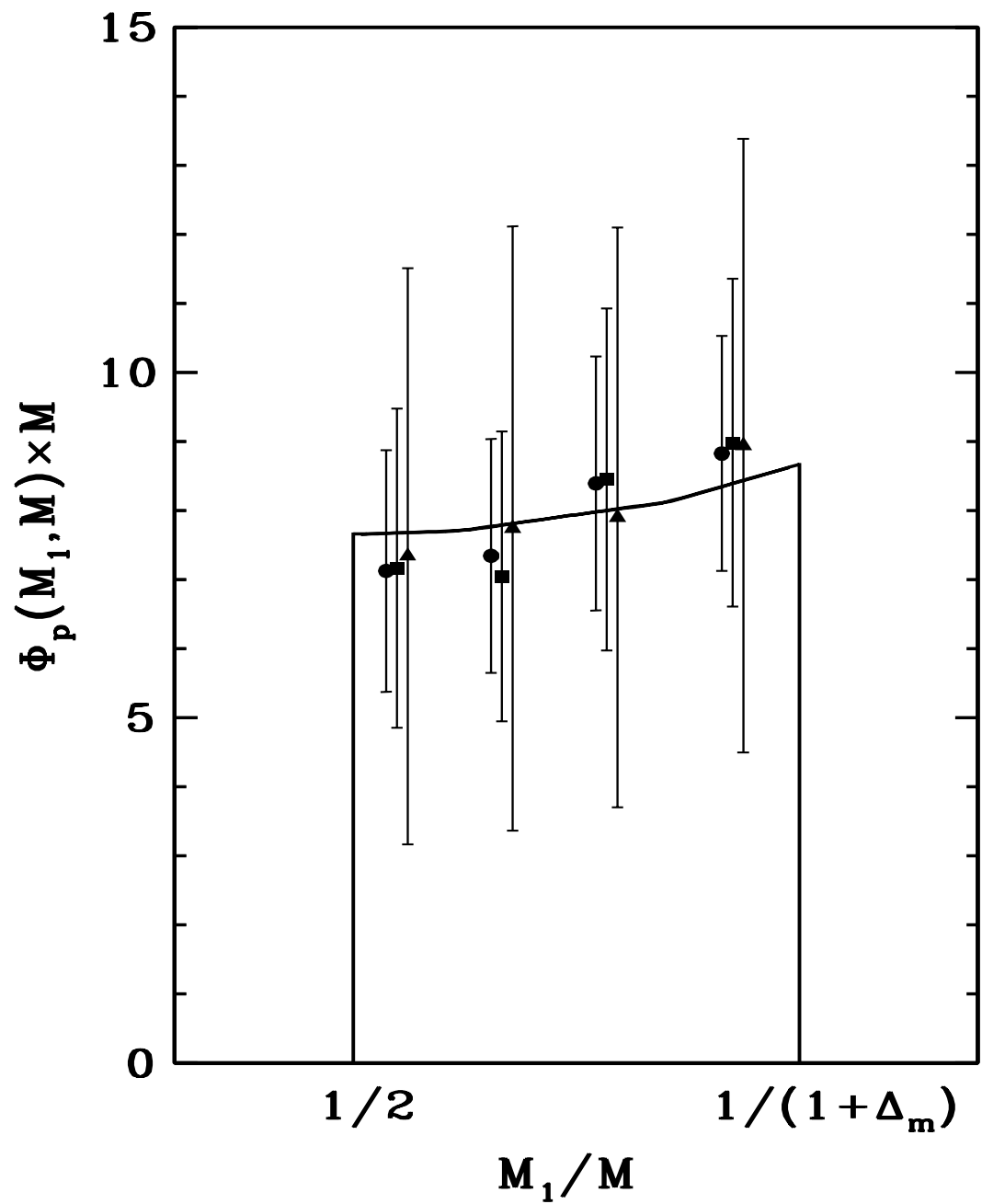,height=21cm,width=\textwidth}
\caption{Comparison between theoretical (solid line) and empirical
(symbols) PDFs of primary progenitor masses $M_1$ for a halo with
arbitrary mass $M$ at formation in the three cosmologies analysed
(triangles for $n=-2$, squares for $n=-1$, and circles for $n=0$). For
clarity, we have shifted the abscissae of the empirical points
corresponding to the $n=-2$ and $n=0$ cosmologies slightly, their true
location coinciding with those of the $n=-1$ cosmology.}
\label{progdi}
\end{figure}

Therefore, given two progenitors at $t$ contributing with $N_1$ and
$N_2$ particles, respectively, to the mass of a halo formed between
$t$ and the next output time, we store the quantity
\begin{equation}
{1\over 1+ N_2/N_1}={N_1\over N_1+N_2}\,,
\end{equation}
which gives an estimate of the fraction of the halo mass at formation
that arises from the primary progenitor at that time. The PDF of this
quantity must be compared with the theoretical PDF of $M_1/M$ ratios,
$\Phi_{\rm p}(M_1/M)$, which coincides with the PDF of primary
progenitor masses $\Phi_{\rm p}(M_1,M)$ multiplied by the mass $M$ of
the halo at formation. This comparison is shown in
Figure~\ref{progdi}; note that this PDF is independent of
time. Once again, the agreement between theory and simulations is very
good.

What is actually plot in Figure~\ref{progdi} deserves some extra
explanation.  From expressions (\ref{mpro}) and (\ref{fung}) it can be
proved that the function $\Phi_{\rm p}(M_1,M) M$ has a very weak
dependence on the spectral index $n$ of the power law spectra of
density fluctuations. Concretely, one has
\begin{equation}
M\,\Phi_{\rm p}(M_1,M)= {2\,x^{(n-9)/6}\over (1-x^{(n+3)/3})^{3/2}}
\left[\int_{\delm/(1+\delm)}^{1/(1+\delm)} {\tilde x^{(n-9)/6}\over 
(1-\tilde x^{(n+3)/3})^{3/2}}
\,\der \tilde x \right]^{-1}\,, 
\label{pwpro}
\end{equation}
with $x=M_1/M$. For any value of $M$, equation (\ref{pwpro}) yields
less than $1\%$ difference among the PDFs of primary progenitor masses
predicted for the three values of $n$ of the cosmologies
considered here. Given the resolution used in Figure~\ref{progdi}, the PDF
of primary progenitor masses predicted by the MPS model for the three
$n=0$, $-1$, and $-2$ cases should overlap on one single curve. This
is the reason why there is only one unique theoretical curve plotted
on this occasion for the three cosmologies. In other words, the \mps
makes a very restrictive prediction which, as seen in Figure
\ref{progdi}, is also fully confirmed by the results of \nbody
simulations for the overall range of primary progenitor masses covered
by the distributions.

\section{SUMMARY AND DISCUSSION}\label{diss}

We have compared the theoretical predictions of the MPS model with the
output of cosmological \nbody simulations in three Einstein-de Sitter
scale-free cosmologies. The comparison included all quantities that
play a fundamental role in the new model and have no counterpart in
the EPS theory, namely, the formation rate, the destruction rate, the
mass accretion rate, the theoretical accretion track, and the PDFs of
halo formation times and progenitor masses. Overall, agreement was
very good, which proves the validity of the MPS model as a powerful
analytic tool for describing the growth history of dark-matter haloes
in hierarchical cosmologies. In addition, the \mps has the important
advantage of avoiding the shortcomings that are encountered when
constructing Monte Carlo merger trees from the usual EPS model.

We emphasize that the agreement between theory and simulations shown
here does not depend on the particular value of $\delm$ used to
establish the effective frontier between minor and major mergers
(provided, of course, that the same value is adopted in both theory
and simulations). The results shown throughout this paper to
illustrate the behaviour of the model correspond to $\delm=0.7$,
although equally good results are obtained for any other value for this
parameter.

As pointed out by Kitayama \& Suto (1996) and SSM (and more recently
by other authors, e.g., RGS; Percival \& Miller 1999; Percival, Miller
\& Peacock 2000; Cohn, Bagla \& White 2001), the distinction between
minor and major mergers is necessary to properly define the concepts
of halo formation and destruction. This is the key element making it
possible to derive, in the \mpsc analytical expressions for the PDFs
of formation times and halo progenitor masses which cannot be
consistently derived in the framework of the EPS model. 

The fact that the exact location of the effective frontier between
minor and major mergers is irrelevant for the agreement between theory
and simulations seems to suggest that the only reason for introducing
a consistent definition for halo formation and destruction is to solve
the problems met by the EPS model when dealing with Monte Carlo merger
trees. Actually, as shown by Manrique el al.~(2001), the formation of
haloes through major mergers and their subsequent evolution through
accretion determine their internal properties. Hence, the proper
characterization of these events is necessary to have a consistent
description of both the mass growth and structure of haloes. A new
model of galaxy formation based on the \mps is currently under
development.

\vspace{0.75cm} \par\noindent 
{\bf ACKNOWLEDGMENTS} \par

\noindent We thank Shaun Cole and Cedric Lacey for sending us 
the data of their $N$-body simulations and, particularly, C. Lacey
for suggesting the comparison between the MPS model and
simulations. This work was supported by the DGI of Spain through
project number AYA2000-0951. A.~Raig, who was supported by an FPI grant
from the MEC of Spain, thanks the staff of TAC in Denmark for their
hospitality during his stay at the beginning of this work.

\newpage
\newpage
\newpage
\newpage
\newpage
\newpage
\newpage

\end{document}